\documentclass[12pt]{article}
\input{epsf}

\makeatletter
\def\Ddots{\mathinner{\mkern1mu\raise\p@
\vbox{\kern7\p@\hbox{.}}\mkern2mu
\raise4\p@\hbox{.}\mkern2mu\raise7\p@\hbox{.}\mkern1mu}}
\makeatother
\font\bm=cmmib10 at 10pt
\font\bms=cmmib10 at 7pt \textfont9=\bm \scriptfont9=\bms
\usepackage{amsfonts}
\usepackage{multirow}
\mathchardef\balpha= "790B
\mathchardef\bbeta= "790C
\mathchardef\bTheta= "7902
\mathchardef\bzeta= "7910
\mathchardef\bOmega= "790A
\mathchardef\bGamma= "7900
\mathchardef\bDelta= "7901
\mathchardef\bPhi= "7908
\mathchardef\bphi= "791E
\mathchardef\bomega= "7921
\mathchardef\bxi= "7918
\mathchardef\bet= "7911
\mathchardef\brho= "791A
\mathchardef\btau= "791C
\mathchardef\bmu= "7916
\mathchardef\bvarpi= "7924

\def \lvec{(\kern-.26em(}
\def\pmb#1{\setbox0=\hbox{#1}
\kern-.025em\copy0\kern-\wd0
\kern.05em\copy0\kern-\wd0
\kern-.025em\raise.0433em\box0 }
\mathchardef\btheta= "7912
\usepackage{amsmath}
\usepackage{amssymb}
\usepackage{authblk}
\usepackage{url}

\begin{document}

\title{Net Zero Averted Temperature Increase}
\author[1] {R. Lindzen }
\author[2] {W. Happer }
\author[3]{ W. A. van Wijngaarden}
\affil[1]{Department of Earth, Atmospheric, and Planetary Sciences,
Massachusetts Institute of Technology, U.S.A}
\affil[2]{Department of Physics, Princeton University, U.S.A}
\affil[3]{Department of Physics and Astronomy, York University, Canada}
\renewcommand\Affilfont{\itshape\small}
\date{\today}
\maketitle

\begin{abstract}
Using feedback-free estimates of the warming by increased atmospheric carbon dioxide (CO$_2$) and observed rates of increase, we estimate that if the United States (U.S.) eliminated net CO$_2$ emissions by the year 2050, this would avert a warming of 0.0084 $^{\circ}$C  (0.015 $^{\circ}$F), which is below our ability to accurately measure.
If the entire world forced net zero CO$_2$ emissions by the year 2050, a warming of only 0.070 $^{\circ}$C (0.13 $^{\circ}$F) would be averted. If one assumes that the warming is a factor of 4 larger because of positive feedbacks, as asserted by the Intergovernmental Panel on Climate Change (IPCC), the warming averted by a net zero U.S. policy would still be very small, 0.034 $^{\circ}$C  (0.061 $^{\circ}$F).  For worldwide net zero emissions by 2050 and the 4-times larger IPCC climate sensitivity, the averted warming would be 0.28 $^{\circ}$C  (0.50 $^{\circ}$F). 
\end{abstract}
\newpage
\section{Introduction}
In this note, we show how to simply estimate the averted  temperature increase $\delta T$ that would result from achieving net zero carbon dioxide emissions in the United States (U.S.) or from worldwide net-zero policies.  Straightforward calculations outlined below show that eliminating U.S. CO$_2$ emissions by the year 2050 would avert a temperature increase of
\begin{equation}
\delta T=0.0084 \hbox{ $^\circ$C}, \label{int2}
\end{equation}
less than a hundredth of a degree centigrade. 

Computer models are not needed to estimate the averted temperature increase (\ref{int2}). It is given to high accuracy by the simple formula
\begin{equation}
\delta T = S\log_2\left(\frac{C}{C'}\right),
\label{int4}
\end{equation}
where $\log_2$ denotes the base-2 logarithm function.  

In (\ref{int4}) the symbol $S$ denotes the equilibrium temperature increase  caused by a doubling of atmospheric  CO$_2$ concentrations.  We will assume a numerical value
\begin{equation}
S=0.75\hbox{ $^\circ$C}.\label{int6}
\end{equation}
Because it is so hard to determine how much of the warming of the past two centuries has been from natural causes and how much is due to increasing concentrations of greenhouse gases, it is not possible to obtain a reliable estimate of $S$ from observations.  The value (\ref{int6}) is a straightforward, feedback-free  estimate that comes from the basic physics of radiation transfer. For example, see p. 19 in the recent review of climate sensitivities \,\cite{Lindzen}. The value (\ref{int6}) is almost the same as the estimate of Rasool and Schneider\,\cite{Schneider}, $S=0.8$~C in the year 1971, before global-warming alarmism became fashionable.

In (\ref{int4}) the symbol $C$ denotes the concentration of atmospheric CO$_2$ in the net-zero target year 2050 if the U.S. takes no measures to reduce emissions. The symbol $C'$ is the concentration  if the U.S. reduces its emissions to zero at that time. 
The U.S. fraction $f_0$ of total world emissions CO$_2$   in the year  2024 is very nearly\cite{U.S._Emissions} 
\begin{equation}
f_0=0.12, \label{int8}
\end{equation}
12\% or about 5 out of 40  billion metric tons of CO$_2$. Most emissions  now are from China and India.  Therefore 
the concentration decrement, $\delta C$, if the U.S. reduces emissions to zero by the year 2050,
\begin{equation}
\delta C=C-C',\label{int10}
\end{equation}
will be relatively small, 
\begin{equation}
\frac{\delta C}{C}\ll 1.\label{int11}
\end{equation}
 We can use (\ref{int11}) to approximate (\ref{int4}) as
\begin{eqnarray}
\delta T 
&=&-S\log_2\left(1-\frac{\delta C}{C}\right)\nonumber\\
&\approx&\frac{S\,\delta C}{\ln(2) \,C}\nonumber\\
&\approx&\frac{S\,f_0 R \Delta t}{2\ln(2)\,(C_0+R\Delta t)}.
\label{int12}
\end{eqnarray}

Before turning to the derivation of (\ref{int12}), which assumes the U.S. fraction of world emissions decreases steadily from $f_0 = 0.12$ now to zero in the year 2050,  we discuss the meanings  of the symbols and we give representative values of them.
The natural (base-$e$) logarithm of $2$,  which appears in (\ref{int12}), has the numerical value
\begin{equation}
\ln (2)=0.6931.\label{int14}
\end{equation}
The atmospheric concentration of CO$_2$ now (the middle of the year 2024) is\,\cite{C}
\begin{equation}
C_0= 427\hbox{ ppm}.\label{int16}
\end{equation}
The time remaining  to the net zero target date of 2050 is
\begin{equation}
\Delta t = 25.5\hbox{ year},\label{int18}
\end{equation}
The current rate of increase of atmospheric concentrations of CO$_2$ is
\begin{equation}
R = 2.5\hbox{ ppm year$^{-1}$}.\label{int20}
\end{equation}
Substituting numerical values from  (\ref{int6}), (\ref{int8}), (\ref{int14}), (\ref{int16}), (\ref{int18})  and (\ref{int20}) into the bottom line of (\ref{int12}) gives (\ref{int2}).

\section{Details}
 If there were no reductions of the U.S. fraction of CO$_2$ emissions, the atmospheric concentration at the net zero target date would be
\begin{eqnarray}
C&=&C_0+\Delta C\nonumber\\
&=&490.75 \hbox{ ppm}.
\label{dt2}
\end{eqnarray}
If the emission rate continues at the constant value $R$ for the time $\Delta t$
the concentration increment would be
\begin{eqnarray}
\Delta C&=&R\Delta t\nonumber\\
& =& 63.75 \hbox{ ppm}.\label{dt4}
\end{eqnarray}
We used (\ref{int18}) and (\ref{int20}) to write the bottom line of (\ref{dt4}), and we used the bottom line of (\ref{dt4}) with (\ref{int16}) to write the bottom line of (\ref{dt2}).
Because the radiative forcing of CO$_2$ is proportional to the logarithm of the concentration, the temperature increment in the year 2050, caused by the concentration increment (\ref{dt4}), would be
\begin{eqnarray}
\Delta T &=&S\log_2\left(\frac{C}{C_0}\right)\nonumber\\
&=&0.1506\hbox{ $^\circ$C} .\label{dt6}
\end{eqnarray}
The numerical values of $S$ from (\ref{int6}), of  $C_0$ from (\ref{int16}) and $C$ from the bottom line of (\ref{dt2}) were used to  evaluate the bottom line of (\ref{dt6}).

The proportionality of the temperature increment $\Delta T$ to the logarithm of the  concentration ratio $C/C_0$ means that the warming from increased CO$_2$ concentrations $C$ is ``saturated.'' That is, each increment $dC$ of CO$_2$ concentration causes less warming than the previous equal increment. Greenhouse warming from CO$_2$ is subject to the law of diminishing returns.

If the U.S. continued to contribute the same fraction $f_0$  of (\ref{int8}) to world CO$_2$ emissions between now and the net zero target date, the U.S. contribution to (\ref{dt4}) would be $f_0 R \Delta t = 7.65$ ppm. But if the U.S. fraction of emissions decreased steadily to zero in the year 2050, the concentration decrement (\ref{int10}) would be
\begin{eqnarray}
\delta C&=&\int_0^{\Delta t} dt\, R f_0\left(1-\frac{t}{\Delta t}\right)\nonumber\\
&=&\frac{1}{2}f_0 R \Delta t\nonumber\\
&=&3.83\hbox{ ppm}.
\label{dt8}
\end{eqnarray}
We used the numerical values of (\ref{int8}) and (\ref{dt4}) to evaluate the bottom line of (\ref{dt8}). Compared to the increase $\Delta T$ of (\ref{dt6}),  the temperature would increase by a slightly smaller amount for a U.S. net zero scenario,
\begin{eqnarray}
\Delta T' &=&S\log_2\left(\frac{C-\delta C}{C_0}\right)\nonumber\\
&=& 0.1421\hbox{  $^\circ$C} .\label{dt10}
\end{eqnarray}
The averted temperature increase $\delta T$ from net-zero policies is 
\begin{eqnarray}
\delta T&=&\Delta T-\Delta T' \nonumber\\
&=&0.0085\hbox{  $^\circ$C}.
\label{dt12}
\end{eqnarray}
The bottom line of (\ref{dt12}) came from subtracting the bottom line of (\ref{dt10}) from the bottom line of (\ref{dt6}).

We can use the top lines of (\ref{dt6})  and (\ref{dt10})  to find a convenient formula for $\delta T$ 
\begin{eqnarray}
\delta T=\Delta T-\Delta T' &=&S\left[\log_2\left(\frac{C}{C_0}\right)-\log_2\left(\frac{C-\delta C}{C_0}\right)\right]\nonumber\\
&=&S\log_2\left(\frac{C}{C-\delta C}\right)\nonumber\\
&=&-S\log_2\left(1-\frac{\delta C}{C}\right).
\label{dt14}
\end{eqnarray}
Recall that the base-2 logarithm, $\log_2(x)$, of some number $x$ is related to the base-$e$ (natural) logarithm, $\ln(x)$, by
\begin{equation}
\log_2(x)=\frac{\ln (x)}{\ln(2)}.\label{dt16}
\end{equation}
Using the power-series expansion
\begin{equation}
-\ln (1-r)=r+\frac{r^2}{2}+\frac{r^3}{3}+\frac{r^4}{4}+\cdots \label{dt18}
\end{equation}
with the last  line of (\ref{dt14}) we find
\begin{eqnarray}
\delta T&=&\frac {S}{\ln(2)}\left[\left(\frac{\delta C}{C}\right)+\frac{1}{2}\left(\frac{\delta C}{C}\right)^2+\frac{1}{3}\left(\frac{\delta C}{C}\right)^3+\cdots \right]\nonumber\\
&\approx&\frac {S}{\ln(2)}\left(\frac{\delta C}{C}\right)\nonumber\\
&\approx&\frac{S\,f_0 R \Delta t}{2\ln(2)\,(C_0+R\Delta t)}.
\label{dt20}
\end{eqnarray}
Because of (\ref{int11}), each term  on the right of the first line of (\ref{dt20}) is at least 100 times smaller than the previous one. So the first term is a good approximation to the sum. The value from the approximate  formula on the second or third line of (\ref{dt20})  only differs by about 1\% from the exact value of $\delta T$, which is given by the sum of the infinite number of terms on first line. Eq. (\ref{dt20}) completes the derivation of (\ref{int12}).
\section{Alternate Assumptions}
Using the last line of  (\ref{int12}), we can see what happens if we use alternate assumptions about the averted temperature increase.  For many years  the United Nations  Intergovernmental Panel on Climate Change (IPCC) asserted that the most likely value of the equilibrium climate sensitivity is four times larger than the feedback-free value (\ref{int6}),
\begin{equation}
S=3.0\hbox{ $^\circ$C}.\label{aa2}
\end{equation}
This assumes a positive feedback that increases the warming by 400\%. According to Le Chatelier's principle, most feedbacks in nature are negative. But if we use the dubious value (\ref{aa2}) in (\ref{int12}) we find that the U.S. net zero scenario  would avert a temperature increase of
\begin{equation}
\delta T= 0.034 \hbox{ $^\circ$C}, \label{aa4}
\end{equation}
less than four hundredth of a degree centigrade. 

As  less developed countries use fossil fuels to raise their standards of living, it is reasonable to expect that the  rate of growth of atmospheric CO$_2$ will increase above the current value, even if the U.S. and other countries implement net zero policies.  Suppose the growth rate increases by  30\%  from the current value of (\ref{int20}) to 
\begin{equation}
R = 3.25\hbox{ ppm year$^{-1}$}.\label{aa6}
\end{equation}
If we use the value (\ref{aa6}) in (\ref{int12}) we find that driving U.S.  CO$_2$ emissions to zero by the year 2050 would avert a temperature increase of
\begin{equation}
\delta T= 0.011 \hbox{ $^\circ$C}, \label{aa8}
\end{equation}
slightly more than one hundredth of a degree centigrade. 

The temperature increment (\ref{aa8}) was estimated for the physically reasonable climate sensitivity $S=0.75\,^{\circ}\hbox{C}$ of (\ref{int6}), and the growth rate $R=3.25$ ppm year$^{-1}$ of (\ref{aa6}) that is 30\% larger than the current growth rate $R=2.5$ ppm year$^{-1}$ of (\ref{int20}). If we use IPCC's  4-times larger, but dubious climate sensitivity  $S=3.0\,^{\circ}\hbox{C}$ of (\ref{aa2}), along with the
larger growth rate $R=3.25$ ppm year$^{-1}$ of (\ref{aa6}), we find an averted temperature increase of
\begin{equation}
\delta T= 0.042 \hbox{ $^\circ$C}, \label{aa10}
\end{equation}
slightly more than four hundredth of a degree centigrade.
\section{Worldwide Net Zero}
We can calculate the averted temperature increase, $\delta T$, 
if the entire world adopted net zero policies and reduced their CO$_2$ emissions to zero by the year 2050. Then the formula for the averted temperature increase would be given by (\ref{int12}) with the fraction $f_0=1$,
\begin{eqnarray}
\delta T 
&=&\frac{S R \Delta t}{2\ln(2)\,(C_0+R\Delta t)}\nonumber\\
&=&0.070\hbox{ $^{\circ}$C}.
\label{wnz2}
\end{eqnarray}
The numerical value of the second line comes from evaluating the expression with the most likely numerical values of (\ref{int6}),
(\ref{int14}), (\ref{int16}), (\ref{int18}) and (\ref{int20}).

Using the four-times larger sensitivity $S= 3\hbox{ $^{\circ}$C}$ of (\ref{aa2}) instead of the more physically reasonable value, $S= 0.75\hbox{ $^{\circ}$C}$ of (\ref{int6}) to evaluate (\ref{wnz2}) we find an averted temperature increase of 
\begin{eqnarray}
\delta T =0.28\hbox{ $^{\circ}$C}.
\label{wnz4}
\end{eqnarray}
\section{The MAGICC Model}
In a prepared statement before the U.S. Senate Budget Committee, B. Zycher \cite{Zycher} showed that the MAGICC model (Model for the Assessment of Greenhouse Gas Induced Climate Change) [6], projects that if the U.S. reduced emissions to zero in the year 2050, the averted temperature increase in the year 2100 would be 
\begin{equation}
\delta T =0.173\hbox{ $^{\circ}$C}.
\label{cnz2}
\end{equation}
The time to net zero for this scenario would be
\begin{equation}
\Delta t = 75.5 \hbox { year},
\label{cnz4}
\end{equation}
instead of $\Delta t = 25.5$ year as in (\ref{int18}). Zycher used an even larger climate sensitivity 
\begin{equation}
S=4.5\hbox{ $^\circ$C},\label{cnz6}
\end{equation}
than the value, $S=3.0\hbox{ $^\circ$C}$ of (\ref{aa2}).
From inspection of (\ref{dt8}) we see  if net US emissions were reduced to zero in a shorter shorter time 
\begin{equation}
\Delta t_{\rm us} = 25.5 \hbox { year},
\label{cnz8}
\end{equation}
than the time $\Delta t=75.5$ years until the year 2100, the averted concentration increment in the year 2100 would be 
\begin{eqnarray}
\delta C&=&R f_0\Delta t -\int_0^{\Delta t_{\rm us}} dt\, R f_0\left(1-\frac{t}{\Delta t_{\rm us}}\right)\nonumber\\
&=& Rf_0\left(\Delta t-\frac{1}{2}\Delta t_{\rm us}\right)\nonumber\\
&=&18.8 \hbox{ ppm}.
\label{cnz10}
\end{eqnarray}
a factor of about 5 larger than (\ref{dt8}) because of the long, 50-year interval from 2050 to 2100 of net zero U.S. emissions. The numerical value on the bottom line of (\ref{cnz10}) was evaluated with (\ref{int8}), (\ref{int18}), (\ref{int20}) and (\ref{cnz8}).

Substituting (\ref{cnz10}) into (\ref{dt20}) we find
\begin{eqnarray}
\delta T
&=&\frac {S}{\ln(2)}\left(\frac{\delta C}{C}\right)\nonumber\\
&=&\frac{S\,f_0 R (2\Delta t-\Delta t_{\rm us})}{2\ln(2)\,(C_0+R\Delta t)}
\nonumber\\&=&0.20 \hbox{ $^\circ$C}.
\label{cnz12}
\end{eqnarray}
The numerical value on the bottom line of (\ref{cnz12}) is reasonably close to the MAGICC estimate (\ref{cnz2}).  It was evaluated with the parameter values from (\ref{int8}),  (\ref{int16}), (\ref{int20}) and (\ref{cnz4}) -- (\ref{cnz8}).
\section{Conclusion}
As shown by (\ref{int2}),  (\ref{aa4}), (\ref{aa8}) and (\ref{aa10}), there appears to be no credible scenario where driving U.S. emissions of CO$_2$ to zero by the year 2050 would avert a temperature increase of more than a few hundredths of a degree centigrade. The immense costs and sacrifices involved would lead to a reduction in warming approximately equal to the measurement uncertainty.  It would be hard to find a better example of a policy of all pain and no gain.

\end{document}